\documentclass[a4paper, 11pt]{article}

\usepackage{a4wide}
\usepackage{latexsym}
\usepackage{graphicx}
\usepackage{color}
\usepackage[pdfstartview=FitH]{hyperref}
\usepackage{verbatim}

\title{A multiplicative process for generating the rank-order distribution of UK election results}

\author{Trevor Fenner, Mark Levene, and George Loizou \\
Department of Computer Science and Information Systems \\
Birkbeck, University of London \\
London WC1E 7HX, U.K. \\ \{mark,trevor,george\}@dcs.bbk.ac.uk}

\date{}
\begin{document}

\maketitle

\begin{abstract}

Human dynamics and sociophysics suggest statistical models that may explain and provide us with a better understanding of social phenomena. Here we propose a generative multiplicative decrease model that gives rise to a rank-order distribution and allows us to analyse the results of the last three UK parliamentary elections. We provide empirical evidence that the additive Weibull distribution, which can be generated from our model, is a close fit to the electoral data, offering a novel interpretation of the recent election results.

\end{abstract}


\section{Introduction}

Recent interest in complex systems, such as social networks, the world-wide-web, email networks and mobile phone networks \cite{BARA07}, has led researchers to investigate the processes that could explain the dynamics of human behaviour within these networks. Human dynamics is not limited to the study of behaviour in communication networks, and has a broader remit similar to the aims of {\em sociophysics} \cite{GALA08,SEN14,STAU14}, which uses concepts and methods from statistical physics to investigate social phenomena and political behaviour. In sociophysics, critical phenomena are important in demonstrating how the transition to global behaviour can emerge from the interactions of many individuals, for example, as described in \cite{PAN10}, where the popularity of movies emerges as collective choice behaviour. The employed methodology often involves the empirical investigation of collective choice dynamics, resulting in the postulation of statistical laws governing, for example, the universal properties of election results \cite{CHAT13}. A fundamental issue in the process of statistical model building is that the model must be tested against experimental data, and superseded by a newer model that better explains the data when such a model is found \cite{GALA08}.

\smallskip

In the context of human dynamics, we have been particularly interested in formulating {\em generative models} in the form of stochastic processes by which complex systems evolve and give rise to power laws or other distributions \cite{FENN05,FENN12,FENN15}. This type of research builds on the early work of Simon \cite{SIMO55}, and the more recent work of Barab\'asi's group \cite{ALBE01} and other researchers \cite{BORN07a}. In the bigger picture, one can view the goal of such research as being similar to that of {\em social mechanisms} \cite{HEDS98}, which looks into the processes, or mechanisms, that can explain observed social phenomena. Using an example given in \cite{SCHE98a}, the growth in the sales of a book can be explained by the well-known logistic growth model \cite{TSOU02}, and more recently we have shown that the process of conference registration with an early bird deadline can be modelled by bi-logistic growth \cite{FENN13}.

\smallskip

In this paper we employ a multiplicative process \cite{REDN90,MITZ04,ZANE08} that turns out to be described by the same underlying equations as the generative model proposed in \cite{FENN15}; the latter captures the essential dynamics of survival analysis applications. Survival analysis \cite{LAWL03,KLEI12} provides statistical methods to estimate the time until an event occurs, known as the {\em survival} or {\em failure} time. Typically, an event in a survival model is referred to as a failure, since it often has negative connotations, such as mortality or the contraction of a disease, although it could be positive, such as the time to return to work or to recover from a disease.

\smallskip

Here we introduce rank-ordering into the model as a natural mechanism in situations where there is no intrinsic ordering of the data, such as constituency-based election results. Rank-ordering statistics \cite{SORN96} is a technique in which we rank the data objects according to size, then plot size against rank, and finally analyse the resulting distribution. Examples of rank-order distributions are: the distribution of large earthquakes \cite{SORN96}, the distribution of oil reserve sizes \cite{LAHE98}, Zipf's rank-frequency distribution \cite[Section~1.4.3]{MANN99}, the size distribution of cities  \cite{BRAK99}, and the distribution of historical extreme events \cite{CHEN12}.

\smallskip

Curtice and Firth \cite{CURT08} and Curtice et al. \cite{CURT11} proposed a pragmatic transformation of the election data to estimate the probability of winning a parliamentary seat in the UK elections. However, they clearly stated in \cite{CURT11} that
``It is not derived from any specific probability distribution for the residuals, but is a purely pragmatic formula which turns estimated shares into numbers between 0 and 1 as required''.  This statement motivated us to investigate the {\em Weibull distribution} \cite{RINN09}, also known as the {\em stretched exponential distribution} \cite{LAHE98}, as a possible fit to the UK elections data set. As a precedent, it is worth noting that Da Paz et al. \cite{DAPA15} proposed a Weibull mixture model to describe the votes obtained by a political party in Brazilian presidential elections.

\smallskip

Although the Weibull distribution does not provide such a good fit for the UK election data of the two main political parties, we found that a variant of the Weibull distribution, known as the {\em additive Weibull distribution} \cite{XIE95,LEMO14,LAI14}, is a suitable model for this purpose. In particular, the additive Weibull is capable of handling {\em bathtub-shaped} ``failure rate functions'' \cite{XIE95,NADA09}, which turn out to be useful for the application to election results.
Thus the main contribution of this paper is to propose the additive Weibull as a distribution for modelling the UK election results within a generative multiplicative model. This proposal is backed up by a thorough analysis of empirical data from the results of the last three UK parliamentary elections.

\medskip

The rest of the paper is organised as follows.
In Section~\ref{sec:urn} we present a multiplicative process that provides us with a mechanism to model the essential dynamics of ranked-ordered models, and derive a well-known differential equation to describe the process. In Section~\ref{sec:fptp} we show how the ``first-past-the-post'' voting system can be modelled using the multiplicative process presented in Section~\ref{sec:urn}, via the rank-ordering method. In Section~\ref{sec:weibull} we introduce the additive Weibull distribution, which is then used in Section~\ref{sec:uk} to model the electoral results of the main parties in the UK over the last three general elections.
Finally, in Section~\ref{sec:conc} we give our concluding remarks.

\section{A multiplicative process for generating a rank-order distribution}
\label{sec:urn}

We next present a generative model in the form of a {\em multiplicative process} \cite{REDN90,MITZ04,ZANE08} that can also be viewed as a survival model similar to the one introduced in \cite{FENN15}, in the context of human dynamics.
In its simplest form, a multiplicative process generates a log-normal distribution \cite[Chapter 14]{JOHN94} \cite{LIMP01}, and has applications in many fields, such as economics, biology and ecology \cite{MITZ04}.
The solution to the multiplicative process we propose will be utilised in Section~\ref{sec:fptp}, in the context of a rank-ordered statistical model of the proportion of votes attained within parliamentary constituencies for a given party in a general election.

\smallskip

We assume a countable number of indices where, for a given party, the $i$th index represents the $i$th constituency ranked in descending order of the number of votes. For any stage $t$, $t \ge 0$, we let $\mu(i,t)$, $0 \le \mu(i,t) \le 1$, be the probability that a potential vote is lost in the $i$th constituency at that stage; $\mu(i,t)$ is known as the {\em mortality rate function} (or simply the mortality function). We always require that $\mu(0,t)=0$ for all $t$.

\smallskip

We now let $F(i,t)$, $0 \le F(i,t) \le 1$, be a discrete function representing, for a given party, the expected proportion of the popular vote potentially attainable by the party in constituency $i$ at stage $t$. Initially, we set $F(0,0) = 1$ for a dummy $0$th constituency, and $F(i,0) = 0$ for all $i > 0$.

\smallskip

The dynamics of the multiplicative process can be captured by the following two equations:
\begin{equation}\label{eq:init}
F(0,t) = 1 \ \ {\rm for} \ \ t \ge 0,
\end{equation}
and
\begin{equation}\label{eq:diff}
F(i+1,t+1) = \left(1 - \mu(i,t)\right) F(i,t) \ \ {\rm for} \ \ 0 \le i \le t.
\end{equation}
\smallskip

Equations (\ref{eq:init}) and (\ref{eq:diff}) define the expected behaviour of a stochastic process \cite{ROSS96} describing how, as $i$ increases, the vote decreases in constituencies where the given party is less popular. For any particular vote, the mortality function is the probabilistic mechanism that decides whether the vote will be lost or not. The process obeys Gibrat's law \cite{EECK04}, which in its original form states that the proportional rate of growth of a firm is independent of its absolute size. In our context, Gibrat's law states that the proportional rate of decrease in the popular vote is independent of the actual number of votes in the constituency.

\medskip

As in \cite{FENN15}, we approximate the discrete function $F(i,t)$ by a continuous function $f(i,t)$, and $\mu(i,t)$ now also a continuous function; $f(i,t)$ is known as the {\em survival function}. Initially, we have $f(0,t)= 1$ for all $t$, and $f(i,0) = 0$ for all $i > 0$.

\smallskip

The dynamics of the model is now captured by the first-order hyperbolic partial differential equation \cite{LAX06},
\begin{equation}\label{eq:partial}
\frac{\partial f(i,t)}{\partial t} + \frac{\partial f(i,t)}{\partial i} + \mu(i,t) f(i,t) = 0,
\end{equation}
which is the same as that used in age-structured models of population dynamics \cite{CHAR94}.

\smallskip

Equation (\ref{eq:partial}) is the well-known {\em transport equation} in fluid dynamics \cite{LAX06}, and
the {\em renewal equation} in population dynamics \cite{PILA91,CHAR94,LI08}.
Following Equation $1.22$ in \cite{CHAR94}, the solution of (\ref{eq:partial}), when $i \le t$, is given by
\begin{equation}\label{eq:renewal}
f(i,t) = \exp \left( - \int_{0}^{i} \mu \left( i-s, t-s \right) ds \right).
\end{equation}

\section{Application of the model to first-past-the-post voting systems}
\label{sec:fptp}

We consider a ``first-past-the-post'' voting system \cite{HORO03}, such as in the UK, where the candidate who receives the most votes in a constituency is elected as the member of parliament for that constituency (or seat). In particular, we will concentrate on the last three UK general elections, which took place in 2005, 2010 and 2015, and the three main parties in the UK at the time, Conservative, Labour and Liberal Democrat. For each party and parliamentary seat, out of a total of 650 seats, we consider the proportion of the popular vote the party attained for that seat.

\smallskip

We make use of the rank-ordering technique \cite{SORN96} in the context of a {\em voter model} as follows.
After fixing the party, we model the proportion of votes attained in seat $i$ by $V_i$, where  $i$ represents the {\em rank} of the seat and $0 \le V_i \le 1$. Thus, ordering the seats in descending order of votes, we obtain the {\em votes vector} $(V_0, V_1, \cdots)$ where:
\begin{equation}\label{eq:votes}
V_0 > V_1 > V_2 > \cdots V_i > \cdots.
\end{equation}
\smallskip

Seat $0$ is a ``dummy'' seat with $V_0 = 1$. In the unlikely event that two seats have the same proportion of votes, their order is chosen randomly.

\smallskip

The votes vector is analogous to the empirical survival function $\hat{S}(\theta)$ \cite{KLEI12}, where $V_i$, which corresponds to $\hat{S}(i)$, can be viewed as an estimate of the expected proportion of the popular vote for seat $i$, given that $V_{i-1}$ was the proportion for seat $i-1$; cf. the {\em Kaplan-Meier estimator} \cite{KAPL58,KLEI12} in the context of survival models. In the context of the voter model, we see that $\hat{S}(i) \approx f(i,t)$.

\smallskip

The rank-ordering of the seats as in (\ref{eq:votes}) can be simulated by the multiplicative process described in Section~\ref{sec:urn}, where $i$ corresponds to the $i$th highest ranked seat. An appropriate mortality function $\mu(i,t)$ is used, which is often decreasing in $i$.
In particular, as we will see in Section~\ref{sec:weibull}, we are generally interested in mortality functions $\mu(i,t) = \mu(i)$ that are independent of $t$. In terms of the voter model, as we consider less popular seats for the given party, i.e. those of lower ``rank'' (remembering that a lower rank is represented by a higher seat number), more votes are lost.

\smallskip

Thus, given a suitable mortality function, as presented in the next section, we can generate the distribution of the votes vector with the stochastic model of Section~\ref{sec:urn}; see \cite{FENN15}. In the next section, we propose the additive Weibull as a possible distribution of the votes vector, and in Section~\ref{sec:uk} we demonstrate its suitability for modelling the UK parliamentary electoral data over the last three general elections.

\section{The additive Weibull distribution}\label{sec:weibull}

The Weibull distribution \cite{RINN09}, also known as the stretched exponential distribution \cite{LAHE98},
is widely used in survival models \cite{KLEI12} and reliability engineering \cite{OCON12}, and it is therefore important to be able to model it. The mortality rate function (also known as the {\em failure rate function}) of the Weibull distribution follows a
power law, which is a monotonic function, and it thus cannot model non-monotonic mortality functions such as the well-known {\em bathtub-shaped} functions \cite{NADA09}.

\smallskip

One solution to the above limitation of the Weibull distribution is to allow its mortality function to be a mixture of two power laws, giving rise to the {\em additive Weibull distribution} \cite{XIE95,LEMO14,LAI14}. Although it is not the only generalisation of the Weibull model allowing bathtub-shaped mortality functions, it has a natural interpretation in terms of a model that can fail in one of two different failure modes \cite{XIE95}. In the first mode the mortality function decreases $-$ this is known as the ``infant mortality'' failure region; in the second mode the mortality function increases $-$ this is known as the ``wear-out'' failure region \cite{NADA09}. Often a third mode, in-between these two, is also identified, where the mortality function does not change rapidly $-$ this is known as the ``random'' failure region. In the context of the stochastic model we have proposed, the additive Weibull distribution is generated when the mortality function is a mixture of two power laws of the form:
\begin{equation}\label{eq:mortality}
\mu(i,t) = \mu(i) = \alpha_1 \beta_1 i^{\beta_1-1} + \alpha_2 \beta_2 i^{\beta_2-1},
\end{equation}
for $i \ge 0$. The {\em scale} parameters $\alpha_1, \alpha_2$ and the {\em shape} parameters $\beta_1, \beta_2$ are all non-negative. Using (\ref{eq:renewal}), we see that the survival function of the additive Weibull distribution has the form:
\begin{equation}\label{eq:additive-weibull}
f(i,t) = f(i) = \exp(- \alpha_1 i^{\beta_1} - \alpha_2 i^{\beta_2}),
\end{equation}
noting that when $\alpha_2 = 0$ it reduces to the standard Weibull survival function.

\smallskip

In particular, bathtub-shaped mortality functions arise when $\beta_1 < 1$ and $\beta_2 > 1$, or $\beta_1 > 1$ and $\beta_2 < 1$, which, as we will see in Section~\ref{sec:uk}, is applicable to the voter model we presented in Section~\ref{sec:fptp}.
We now assume, without loss of generality, that $\beta_1 < 1$ and $\beta_2 > 1$. Then, as was shown in \cite{XIE95,LEMO14}, the {\em change point} $i^*$ is the value of $i$ for which
\begin{displaymath}
\frac{d \mu(i)}{d i} = 0.
\end{displaymath}
\smallskip

This has the unique solution
\begin{equation}\label{eq:change-point}
i^* = \left[ \frac{\alpha_1 \beta_1 (1- \beta_1)}{\alpha_2 \beta_2 (\beta_2 -1)} \right]^{\frac{1}{\beta_2 - \beta_1}}.
\end{equation}
\smallskip

It is straightforward to show that $\mu(i)$ is always a minimum at $i=i^*$.
So, for $i < i^*$ the mortality function (\ref{eq:mortality}) is decreasing and for $i > i^*$ it is increasing, implying that the mortality function is bathtub-shaped. Now $\mu(i)$ is strictly increasing for $i > i^*$, but must be bounded above to satisfy $0 \le \mu(i) \le 1$ (cf. \cite{STEI84}). This does not pose a problem in our application, as the number of parliamentary seats in the UK is $650$ and, as we shall see in Section~\ref{sec:uk}, $\mu(i) \le 1$ when $i \le 650$.

\section{Analysis of the last three UK parliamentary elections}
\label{sec:uk}

We now analyse the performance of the three major parties, Conservative, Labour and Liberal Democrat in the last three UK parliamentary elections in 2005, 2010 and 2015; the electoral data is available online at \url{www.electoralcommission.org.uk/our-work/our-research/electoral-data}.

\smallskip

We will make use of the {\em Jensen-Shannon divergence} ({\em JSD}) \cite{ENDR03}, which is a nonparametric measure of the distance between two distributions $\{p_i\}$ and $\{q_i\}$, where $i=1,2,\ldots, n$. The formal definition of the {\em JSD}, which is a symmetric version of the Kullback-Leibler divergence and is based on Shannon's entropy \cite{COVE91}, is given by
\begin{equation}\label{eq:jsd}
JSD(\{p_i\},\{q_i\}) = \sqrt{\frac{1}{2 \ln{2}} \ \sum_{i=1}^{n} \left( p_i \ \ln{\frac{2 p_i}{p_i + q_i}} + q_i \ \ln{\frac{2 q_i}{p_i + q_i}} \right)},
\end{equation}
where we use the convention that if $p_i = 0$ or $q_i = 0$, or both, the corresponding term in the summation is defined to be $0$;
the factor $2 \ln{2}$ is included to normalise the {\em JSD}. We note that either $\{p_i\}$ or $\{q_i\}$ may be incomplete, in which case either $\sum_i p_i$ or  $\sum_i q_i$ would be strictly less than one. In this case we may replace the normalisation factor $2 \ln{2}$ in (\ref{eq:jsd}) by $\sum_i (p_i+q_i) \ln{2}$.

\smallskip

We next outline the methodology we have used to validate and evaluate the model presented in Section~\ref{sec:fptp}, noting that all computations were carried out using the Matlab software package.

\renewcommand{\labelenumi}{(\roman{enumi})}
\begin{enumerate}

\item We first download the electoral data and preprocess it to obtain the {\em raw data} in the form of vote vectors satisfying (\ref{eq:votes}), for the three parties for each election.

\item For each party and each election, we then perform nonlinear regression to fit an additive Weibull survival function (\ref{eq:additive-weibull}) to the {\em raw data} to obtain the {\em fitted parameters} $\alpha_1, \alpha_2, \beta_1$ and $\beta_2$.  The additive Weibull survival function is fitted to the Conservatives and the Labour electoral data, while the standard Weibull survival function (with $\alpha_2 = 0$) is fitted to the Liberal Democrat electoral data.

    We note that the standard Weibull survival function does not exhibit a good fit to the Conservative and Labour data. On the other hand, while the additive Weibull is a good fit to the Liberal Democrat data, as it does not have a bathtub-shaped mortality function since $\beta_1$ and $\beta_2$ are both less than $1$, we can still obtain a good fit using the standard Weibull survival function (which has fewer parameters).
\item For each party, we compute the JSD between the fitted distributions over the three elections, and also between the parties for each election year. This allows us to examine the changes that occurred over the three elections within the parties and between them.
\end{enumerate}
\smallskip

The fitted parameters  $\alpha_1$, $\beta_1$, $\alpha_2$ and $\beta_2$ for the Conservative party are shown in the rows of Table~\ref{table:con}, together with the coefficient of determination $R^2$ \cite{MOTU95}, the mortality probability $\mu(650)$ for the lowest ranked constituency, the change point $i^*$ and $f(i^*)$.
Similarly, the corresponding results for the Labour party are shown in the rows of Table~\ref{table:lab}.
We observe that $\mu(650) < 1$ in all cases, as required.
We note that $i^*$ for the Conservative party is significantly less than for Labour, which may be becuase
Labour attains higher margins in its ``safe'' seats than the Conservatives,
since for values higher than $i^*$ the mortality function increases at a fast rate.
On the other hand, we see that for the Conservatives $f(i^*)$ at the change point has increased to about 53\% from 2005 to 2015, but for Labour has decreased to about 46\%. It is worth noting that the shape parameter $\beta_1$ for Labour in 2015 stands out in its increase from the previous elections. This indicates that for its ``safe'' seats the drop in the vote in 2015 was steeper than in previous elections.

\smallskip

The fitted parameters  $\alpha_1$ and $\beta_1$ for the Liberal Democrat party are shown in the rows of Table~\ref{table:lib}, together with the coefficient of determination $R^2$ \cite{MOTU95} and the mortality probability $\mu(650)$; in this case $\mu(650)$ is only shown for completeness, as it is always between 0 and 1 since $0 < \beta_1 < 1$. We can see that both $\alpha_1$ and $\beta_1$ increased in 2015, accounting for their demise in the last election, while in 2005 and 2010 their values were very similar. Finally, we observe that for all regressions, the $R^2$ values indicate a very good fit for all of the years.

\begin{table}[ht]
\begin{center}
\begin{tabular}{|l|c|c|c|c|c|c|c|c|}\hline
Conservatives & $\alpha_1$ & $\beta_1$ & $\alpha_2$ & $\beta_2$ & $R^2$  & $\mu(650)$ & $i^*$  & $f(i^*)$ \\ \hline \hline
2005          & 0.4551     & 0.1072    & 9.0174e-08 & 2.5951    & 0.9969 & 0.0073    & 108.83 & 0.4631   \\ \hline
2010          & 0.4101     & 0.1044    & 7.4650e-08 & 2.5997    & 0.9954 & 0.0063    & 109.79 & 0.5042   \\ \hline
2015          & 0.3639     & 0.1134    & 3.2490e-08 & 2.7580    & 0.9929 & 0.0080    & 106.90 & 0.5321   \\ \hline
\end{tabular}
\end{center}
\caption{\label{table:con} Nonlinear least-squares regression fitting an additive Weibull survival function to the empirical survival function for the Conservative party election results.}
\end{table}

\begin{table}[ht]
\begin{center}
\begin{tabular}{|l|c|c|c|c|c|c|c|c|}\hline
Labour   & $\alpha_1$ & $\beta_1$ & $\alpha_2$ & $\beta_2$ & $R^2$  & $\mu(650)$ & $i^*$  & $f(i^*)$ \\ \hline \hline
2005     & 0.2742     & 0.1817    & 2.4678e-08  & 2.7462   & 0.9871 & 0.0058     & 144.29 & 0.4977     \\ \hline
2010     & 0.3188     & 0.1779    & 5.3946e-08  & 2.6856   & 0.9964 & 0.0083     & 127.61 & 0.4585     \\ \hline
2015     & 0.2089     & 0.2545    & 8.3618e-08  & 2.5927   & 0.9990 & 0.0070     & 145.87 & 0.4600     \\ \hline
\end{tabular}
\end{center}
\caption{\label{table:lab} Nonlinear least-squares regression fitting an additive Weibull survival function to the empirical survival function for the Labour party election results.}
\end{table}

\begin{table}[ht]
\begin{center}
\begin{tabular}{|l|c|c|c|c|}\hline
Liberal Democarats & $\alpha_1$ & $\beta_1$ & $R^2$ & $\mu(650)$ \\ \hline \hline
2005               & 0.2787     & 0.3114    & 0.9666 & 0.0010 \\ \hline
2010               & 0.2717     & 0.3101    & 0.9636 & 0.0010 \\ \hline
2015               & 0.4362     & 0.3355    & 0.9511 & 0.0020 \\ \hline
\end{tabular}
\end{center}
\caption{\label{table:lib} Nonlinear least-squares regression fitting a Weibull survival function to the empirical survival function for the Liberal Democrat party election results.}
\end{table}
\smallskip

The JSDs for each party between the distributions for the three election years is shown in the rows of Table~\ref{table:party-year}, while the JSDs for each election year between the distributions of the major parties is shown in the rows of Table~\ref{table:party-party}.

\begin{table}[ht]
\begin{center}
\begin{tabular}{|l|c|c|c|}\hline
Party & 2005 vs 2010 & 2010 vs 2015 & 2005 vs 2015  \\ \hline \hline
Con   & 0.0339       & 0.0451       & 0.0575 \\ \hline
Lab   & 0.0438       & 0.0689       & 0.1020 \\ \hline
Lib   & 0.0068       & 0.1530       & 0.1468 \\ \hline
\end{tabular}
\end{center}
\caption{\label{table:party-year} JSDs for each party between the distributions for the different election years.}
\end{table}

\begin{table}[ht]
\begin{center}
\begin{tabular}{|l|c|c|c|}\hline
Party      & 2005   & 2010   & 2015  \\ \hline \hline
Con vs Lab & 0.0908 & 0.1298 & 0.1856 \\ \hline
Con vs Lib & 0.4056 & 0.4246 & 0.5498 \\ \hline
Lab vs Lib & 0.3483 & 0.3255 & 0.4106 \\ \hline
\end{tabular}
\end{center}
\caption{\label{table:party-party} JSDs for each election year between the distributions of the different parties.}
\end{table}
\smallskip

In order to interpret the results, it is worth noting that the most important changes in the 2015 general election that led to the Conservative victory were (see \cite{DENV15}):
\renewcommand{\labelenumi}{(\roman{enumi})}
\begin{enumerate}
\item The demise of the Liberal Democrats, losing most of their seats, with half of them going to the Conservatives; and

\item Labour losing all of their seats but one in Scotland.
\end{enumerate}

Point (i) above is reflected in the second and third rows of Table~\ref{table:party-party}, where we can see that the distance between the Conservatives and the Liberal Democrats was growing, culminating in 54.98\% in 2015, which is 13.92\% more than the difference between Labour and the Liberal Democrats in 2015. In addition, we can see from the first row in Table~\ref{table:party-party} that the distance between the Conservatives and Labour has
grown in 2015 to 18.56\%. Moreover, from the middle column of Table~\ref{table:party-year} we can see that the Conservative distribution changed by 4.51\% from 2010 to 2015, reflecting the additional seats gained in 2015, while the Labour distribution changed by 6.89\% from 2010 to 2015, reflecting their losses in Scotland.

\section{Concluding remarks}
\label{sec:conc}

We have proposed a multiplicative process that generates the rank-order distributions of UK election results. In our model, the proportion of votes attained for a given party decreases according to a specified mortality function, leading to a rank-order distribution represented by the votes vector.
A solution to the continuous approximation of the equation specifying the model was given in (\ref{eq:renewal}), and is therefore identical to that of the renewal equation in population dynamics \cite{CHAR94}.

\smallskip

We then considered how the UK election results using the rank-ordering technique \cite{SORN96}, ordering the seats according to the proportion of votes attained, resulting in the votes vector (\ref{eq:votes}). The votes vector is analogous to the empirical survival function \cite{KLEI12}, and is used to estimate the expected share of the popular vote for seat $i$ in the rank ordering. We suggested the additive Weibull distribution, with its bathtub mortality function (\ref{eq:mortality}), being a mixture of power laws, as a suitable model for the UK election results. This was validated for the last three parliamentary election results in the UK, where both the Conservative and Labour electoral data exhibited a good fit to the additive Weibull, while for the smaller Liberal Democrat party a standard Weibull was sufficient for a good fit to the data. The additive Weibull gives us the extra flexibility to model ``safe'' and ``marginal'' seats for the two major political parties in the UK.
In our analysis we offered an interpretation of the 2015 election victory for the Conservatives by utilising the JSD (\ref{eq:jsd}) to measure the distance between the distributions for the parties over the last three elections.
We stress that the only data we have used in our analysis are the raw counts of votes for the seats from the election results; no demographic or other historical constituency data was used.

\smallskip

Generative models such as the one we have proposed for rank-order distributions can be useful in explaining social phenomena such as voter behaviour in election results. It remains to be seen how the insight gained from such models can inform a predictive regression model, such as conditional logistic regression \cite{ALVA98}, which was used in \cite{CURT11}.

\section*{Acknowledgements}
We would like to thank Muawya Eldaw, who preprocessed the election data sets.

\newcommand{\etalchar}[1]{$^{#1}$}

\end{document}